# WaveNet-SF: A Hybrid Network for Retinal Disease Detection Based on Wavelet Transform in Spatial-Frequency Domain

Jilan Cheng[1], Guoli Long[1], Zeyu Zhang[1], Zhenjia Qi[1], Hanyu Wang[2], Libin Lu[3], Shuihua Wang[4], Yudong Zhang[5], Jin Hong[1,*]

1. School of Information Engineering, Nanchang University, Nanchang, 330031, China
2. School of Advanced Energy, Sun Yat-sen University, Shenzhen, 518107, China
3. School of Mathematics and Computer Science, Wuhan Polytechnic University, Wuhan, 430023, China
4. Department of Biological Sciences, School of Science, Xi'an Jiaotong Liverpool University, Suzhou, 215123, China
5. School of Computer Science and Engineering, Southeast University, Nanjing, 210096, China

E-mail: 6003122043@email.ncu.edu.cn; 6003122121@email.ncu.edu.cn; 5801122002@email.ncu.edu.cn; 6108122008@email.ncu.edu.cn; wanghy553@mail.sysu.edu.cn; rubinius@whpu.edu.cn; Shuihua.Wang@xjtlu.edu.cn; yudongzhang@ieee.org; hongjin@ncu.edu.cn;

* Correspondence should be addressed to Jin Hong

**Abstract:** Retinal diseases are a leading cause of vision impairment and blindness, with timely diagnosis being critical for effective treatment. Optical Coherence Tomography (OCT) has become a standard imaging modality for retinal disease diagnosis, but OCT images often suffer from issues such as speckle noise, complex lesion shapes, and varying lesion sizes, making interpretation challenging. In this paper, we propose a novel model, WaveNet-SF, to enhance retinal disease detection by integrating the spatial-domain and frequency-domain learning. The framework utilizes wavelet transforms to decompose OCT images into low- and high-frequency components, enabling the model to extract both global structural features and fine-grained details. To improve lesion detection, we introduce a Multi-Scale Wavelet Spatial Attention (MSW-SA) module, which enhances the model's focus on regions of interest at multiple scales. Additionally, a High-Frequency Feature Compensation (HFFC) block is incorporated to recover edge information lost during wavelet decomposition, suppress noise, and preserve fine details crucial for lesion detection. Our approach achieves state-of-the-art (SOTA) classification accuracies of 97.82% and 99.58% on the OCT-C8 and OCT2017 datasets, respectively, surpassing existing methods. These results demonstrate the efficacy of WaveNet-SF in addressing the challenges of OCT image analysis and its potential as a powerful tool for retinal disease diagnosis.

## 1. Introduction

The eye is a crucial organ for the human body, responsible for acquiring and processing visual information from the external world. However, visual impairment is becoming a significant global health issue. According to the World Health Organization, by 2050, over 1.7 billion people will be expected to suffer from varying degrees of visual impairment [1]. Retinal diseases, in particular, are a leading cause of vision loss and blindness. Some of the most common retinal conditions include age-related macular degeneration (AMD), choroidal neovascularization (CNV), central serous retinopathy (CSR), diabetic macular edema (DME), yellow deposits under the retina (DRUSEN), and macular hole (MH) [2-5].

These diseases can cause significant damage to vision if not diagnosed and treated early. Timely detection and accurate diagnosis are essential for preventing further deterioration and preserving vision.

Optical Coherence Tomography (OCT) has emerged as a leading non-invasive imaging modality for retinal diagnosis. OCT provides high-resolution cross-sectional images of the retina, offering valuable insights into the structure and condition of retinal layers. This detailed imaging capability allows clinicians to detect and monitor a wide range of retinal and macular diseases [6, 7]. However, interpreting OCT images requires specialized knowledge and extensive diagnostic experience, and the process can be time-consuming. This places a strain on healthcare resources, emphasizing the need for automated systems that can assist clinicians in efficiently and accurately diagnosing retinal diseases.

Deep learning techniques, which are a subset of machine learning, have shown great promise in medical image analysis. By leveraging large datasets and advanced algorithms, deep learning models can automatically learn features from images and perform complex tasks such as disease detection [8, 9], segmentation [10-12] and classification [13, 14]. In retinal disease diagnosis, convolutional neural networks (CNNs) have become the go-to architecture for OCT image analysis due to their ability to extract hierarchical features and learn spatial patterns from images [15]. Several studies have employed CNN-based models to classify retinal diseases, achieving impressive results [15-18]. However, OCT retinal images present unique challenges that hinder the performance of many existing models. These challenges include the presence of lesions with varying shapes, sizes, and locations, as well as the substantial speckle noise that is inherent in OCT images, as shown in **Fig. 1**.

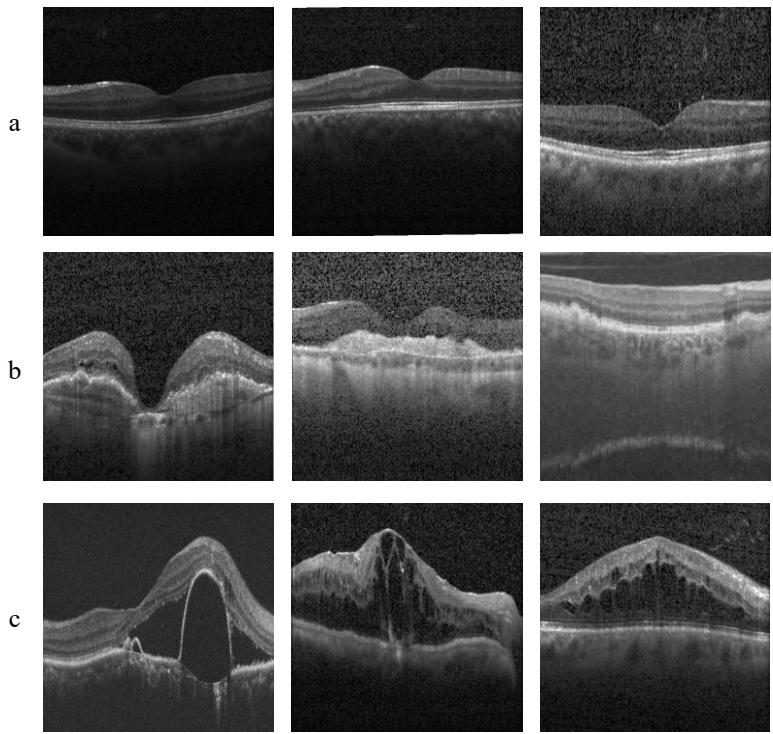

**Fig. 1** Example OCT retinal images. (a) shows a normal retinal image, while (b) and (c) are images of diseased retinas, respectively. In (b), significant speckle noise can be observed in the OCT retinal image, while in (c), the retinal boundary is noticeably different from that of a normal retina.

While deep learning models, such as CNNs, have made significant strides in improving the accuracy of OCT image classification [15-18], they still face limitations. Additionally, the use of multi-scale feature extraction methods, such as Feature Pyramid Networks (FPNs), has become a common strategy to capture information from multiple scales [18, 19]. However, these methods often fail to effectively capture edge information, which is essential for accurately identifying lesions in OCT images. More critically, speckle noise often overlaps and mixes with discriminative features, which not only undermines the model's ability to extract effective representations but also adversely affects the computation of attention mechanisms.

To address the aforementioned challenges, we propose a novel OCT retinal disease detection model, termed WaveNet-SF. This approach integrates frequency-domain and spatial-domain learning: on one hand, it effectively enhances the accuracy of the attention mechanism when processing noisy images and diseases with diverse morphologies, sizes, and locations; on the other hand, it leverages frequency-domain learning to extract and reinforce retinal edge features that are difficult to capture in the spatial domain, thereby significantly improving the model's ability to detect different types of diseases.

Specifically, we propose a novel architecture comprising two branches: a low-frequency feature extraction branch and a high-frequency feature compensation branch. Using wavelet transform, OCT images are decomposed into low- and high-frequency components, where the low-frequency components preserve global structural information and facilitate denoising, while the high-frequency components contain critical edge information, which is essential for detecting subtle lesions. The low-frequency feature extraction branch serves as the primary feature extractor, capturing global structural information from the low-frequency components, whereas the high-frequency feature compensation branch extracts and enhances retinal edges and other fine-grained information beneficial for disease detection, serving an auxiliary role. In addition, we introduce the MSW-SA and HFFC modules.

The MSW-SA module is designed to address the limitations of conventional multi-scale spatial attention mechanisms. Traditional methods typically rely on large-kernel convolutions to achieve a wider receptive field, which requires substantial padding and can reduce the accuracy of attention computation. Moreover, the inherent speckle noise in OCT images can interfere with the spatial attention mechanism's ability to focus on critical regions. It is important to note that the primary purpose of spatial attention is to enhance the model's focus on key structures within the image. Inspired by this, we incorporate wavelet transform into the spatial attention mechanism and use the low-frequency components obtained from wavelet decomposition—which contain the global structural information of the image—for spatial perception and attention computation. Furthermore, we incorporate wavelet decomposition into the spatial attention computation, obtaining low-frequency components at multiple levels and recursively utilizing these components in the attention calculation, thereby enhancing the modeling capability of spatial attention for low-frequency spatial structures. This strategy not only suppresses the interference of high-frequency noise in attention calculation, thereby improving attention accuracy, but also, due to the intrinsic downsampling property of the wavelet transform, effectively enlarges the receptive field, enabling efficient modeling of global structural information. Furthermore, the HFFC module employs a frequency-domain learning approach to selectively enhance different frequency components, thereby extracting and reinforcing retinal edge features as well as other fine-grained features beneficial for disease detection, ultimately improving the discriminative capability of the model. The contributions of this work are as follows:

(i) We propose a hybrid learning model that integrates frequency-domain learning and spatial-domain learning, which successfully extends frequency-domain learning from fine-grained tasks such as image enhancement to the coarse-grained task of retinal fundus classification. This framework captures structural features of the image through spatial-domain learning, and enhances and extracts edge details through frequency-domain learning, thereby effectively improving the discriminative capability of the model. Our code is publicly available on https://github.com/liuliutaotao64/WaveNet-SF

(ii) We design a Multi-Scale Wavelet Spatial Attention (MSW-SA) module that utilizes a multi-scale receptive field to more effectively capture spatial information in OCT images. By incorporating global contextual information through wavelet transform, our method alleviates the interference of noise in spatial attention computation and enhances the modeling of spatial structures, thereby allowing the model to better focus on complex lesion regions.

(iii) We propose a High-Frequency Feature Compensation (HFFC) module, which enhances and extracts edge information and other important features from the high-frequency components that are often buried in noise. Through frequency-domain learning, this module enables the model to capture fine-grained details in the image, thereby improving lesion classification performance under noisy conditions.

(iv) We propose a Network Architecture that separately extracts features from both the low-frequency and high-frequency components of the OCT image. This design enables our model to effectively leverage both global structural and fine-grained detail information, enhancing the classification performance.

(v) Our model achieves superior performance on benchmark datasets, demonstrating state-of-the-art (SOTA) classification accuracies of 97.82% and 99.58% on the OCT-C8 and OCT2017 datasets, respectively. This outperforms existing methods, showing the effectiveness of our approach in addressing the challenges of OCT image classification.

## 2. Related work

### 2.1. OCT Retinal Image Classification

Deep learning has significantly advanced OCT retinal image classification in recent years. Pre-trained networks such as AlexNet, ResNet18, and GoogLeNet have been utilized for automatic detection of retinal conditions like CSR from enhanced and denoised OCT images [20]. Ensemble models based on improved ResNet50 architectures have demonstrated high classification accuracy by integrating predictions from multiple models [21, 22]. Huang et al. [23] introduced the layer-guided convolutional neural network (LGCNN), which utilizes pre-trained ReLayNet to segment specific retinal layers (ILM-RPE and RPE-BrM) and employs parallel subnetworks for feature extraction, achieving superior performance.

The emergence of Transformer-based architectures marked a turning point in OCT image classification. Unlike CNNs, which excel at extracting local spatial features, Transformers capture global dependencies within images [24]. However, pure Transformer models still face challenges in capturing local features and require large amounts of training data. To address these limitations, various hybrid architectures have been proposed. For instance, HTC-retinal [15] integrates Vision Transformer (ViT) and CNN to leverage both global and local contexts, achieving 97.00% accuracy on the OCT-C8 dataset. Similarly, parallel designs combining ConvNet and Transformer branches have enhanced feature fusion, showcasing the strengths of both approaches [25]. Nevertheless, these methods still encounter challenges such as high model complexity and substantial computational resource demands. To this end, Swin-Poly Transformer [16] introduces an adaptive connection mechanism between image regions, improving local information modeling and model interpretability. It achieves an accuracy of 97.11% on OCT-C8, demonstrating superior performance.

Some studies focus on guiding models to pay more attention to lesion-relevant regions. For example, the "Edgen" block replaces traditional ResNet residual blocks to improve sensitivity to retinal boundaries [17]. Fang et al. [26] developed a lesion-aware network (LDCNN) that used soft attention maps to emphasize lesion areas. However, these methods often overlook multi-scale and frequency-domain information, which can enhance feature representation.

Efforts to incorporate multi-scale feature extraction into OCT image classification have shown promise. Iterative Feature Fusion methods [27] and Feature Pyramid Networks (FPN) [19] facilitate hierarchical aggregation of spatial features at different scales, effectively capturing structural information at multiple resolutions. Building upon this, Peng et al. [18] introduced a multi-scale denoising residual network that not only fused features across scales but also mitigated noise interference, thereby improving classification performance. Similarly, Zuo et al. [28] employed the Mamba framework to extract comprehensive multi-scale global features, enhancing the model's representation capacity. Despite these advancements, most methods focus exclusively on spatial-domain features, leaving frequency-domain information underexplored.

**2.2. Spatial Attention Mechanism**

Spatial attention mechanisms have become a cornerstone in computer vision tasks, enhancing model performance by emphasizing critical regions. The Convolutional Block Attention Module (CBAM) introduces a 7×7 large convolution kernel to capture spatial information and generate attention maps [29]. Similarly, the Bottleneck Attention Module (BAM) uses dilation convolutions to expand receptive fields, effectively capturing spatial dependencies [30]. Advanced techniques such as non-local spatial attention modules (NL-SAM) [31] and parameter-free Spatial Intersection Attention Modules (SIAM) [32] adaptively recalibrate feature maps to highlight important regions.

Our proposed Multi-Scale Wavelet Spatial Attention (MSW-SA) module diverges from existing approaches in two key aspects: i) It can effectively mitigate the interference of noise on spatial attention computation. ii) It integrates global contextual information without introducing additional parameters or padding, thereby enhancing the model's spatial modeling capability and improving computational efficiency and attention accuracy.

**2.3. Frequency-Domain Learning**

Traditional CNNs primarily focus on spatial feature extraction, but frequency-domain learning has recently gained traction due to its ability to capture finer-grained details. Frequency-domain learning is to decompose an image into different frequency subbands, and then analyze and process the feature distribution of each frequency subband, so as to enhance the model's ability to model information of different frequencies. The Frequency Selection Network (FSNet) [33] and Multi-Scale Frequency Selection Network (MSFSNet) [34] dynamically decompose features into frequency subbands, enhancing tasks such as image restoration. Inspired by the convolution theorem, FFTformer [35] substitutes the matrix multiplication operation in the spatial domain with element-wise multiplication in the frequency domain, thereby introducing an efficient frequency-domain self-attention module (FSAS). This approach has demonstrated notable effectiveness in the task of image deblurring.

Wavelet transform [36], a classic signal processing technique, has been adapted for frequency-domain learning in deep networks. For instance, replacing standard downsampling with wavelet-based modules preserves semantic information, improving segmentation performance [37]. Similarly, wavelet convolution has been shown to efficiently capture large receptive fields while avoiding excessive parameters [38]. Y-net [39] proposed a wavelet-based structural similarity loss function (W-SSIM), which effectively enhances dehazing performance by aggregating the structural similarity of multi-scale and multi-frequency patches with weighted accumulation.

It is worth noting that, although inspired by FSNet and MSFSNet, our work fundamentally differs from these methods. Both FSNet and MSFSNet were developed for image restoration tasks, aiming to recover sharp images from blurred inputs, which constitutes a fine-grained, low-level vision task. Both methods are based on the U-Net architecture and incorporate frequency-domain learning during the upsampling and downsampling processes to enhance the model's ability to capture image details. In contrast, OCT classification is a coarse-grained task that emphasizes semantic understanding. Previous studies have rarely explored how classification tasks can benefit from subtle image details, particularly in OCT retinal images, where the retinal edges vary significantly across disease categories, and spatial-domain learning alone struggles to capture these fine structures. To address this limitation, we integrate frequency-domain learning into OCT retinal image classification, enabling precise discrimination of subtle inter-class differences.

## 3. Method

### 3.1. Overall architecture

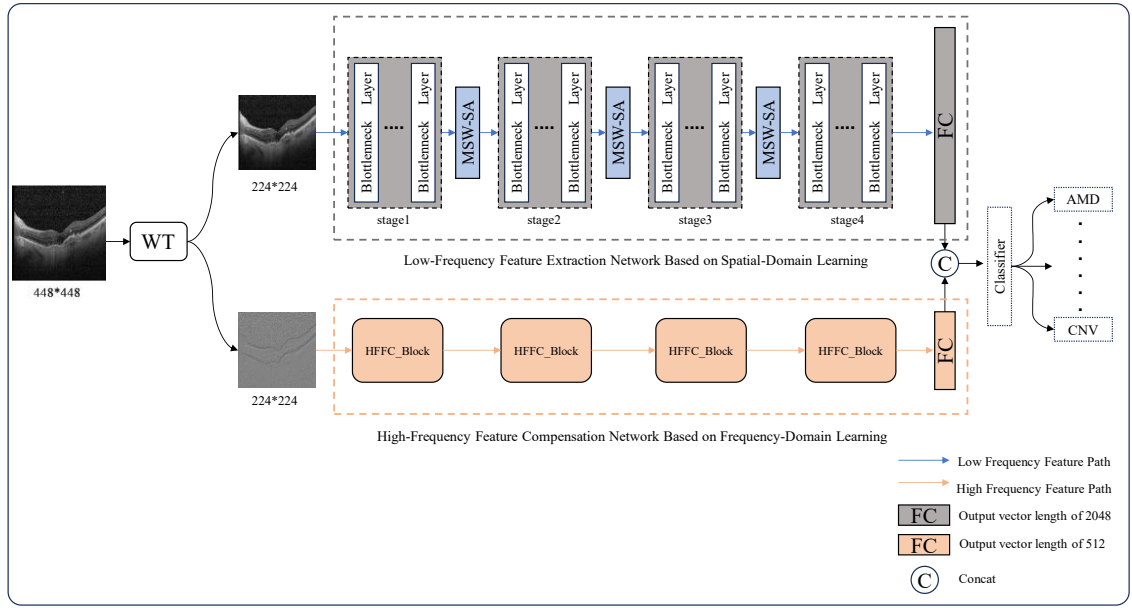

**Fig. 2** The overall framework of WaveNet-SF.

The overall framework of WaveNet-SF is illustrated in **Fig. 2**. The architecture comprises three key components: a wavelet-transform block, a low-frequency feature extraction network, and a high-frequency feature compensation network. The network processes the input OCT retinal image in a structured manner, facilitating effective feature extraction and classification.

Initially, the OCT retinal image undergoes a wavelet transform, which decomposes it into four subbands: LL, HL, HH, and LH. The LL subband, representing the low-frequency approximation of the image, is directed into the low-frequency feature extraction network. This network utilizes ResNet50 as its backbone, chosen for its established efficacy in OCT retinal image classification. To enhance its performance, the proposed MSW-SA module is integrated between the stages of ResNet50, enriching the spatial and contextual representation of the low-frequency features. The network outputs a 2048-dimensional feature vector through its fully connected layer, encapsulating the critical low-frequency information for classification.

Simultaneously, the sum of the HL, HH, and LH subbands is computed to represent the high-frequency components of the image. These components, containing edge details and noise, are input into the high-frequency feature compensation network. This network is structured with four HFFC modules, each designed to extract and emphasize the high-frequency features that are beneficial for classification while suppressing irrelevant noise. The output of this network is a 512-dimensional feature vector, which complements the low-frequency features by providing additional discriminative power.

Finally, the outputs of the low-frequency and high-frequency networks are fused to form the final feature representation, enabling robust and precise classification. This systematic flow from input image to fused feature vector ensures that both coarse and fine-grained information is effectively captured and utilized for accurate analysis.

### 3.2. Wavelet Transform

Wavelet Transform is an effective method for analyzing the frequency domain information of images by decomposing them into localized frequency components. It achieves this by convolving the input image with wavelet basis functions at various scales $j$ and positions $k$, effectively separating the image into different frequency bands. This process results in four subbands: a low-frequency subband (LL) representing the approximation of the image, and three high-frequency subbands (HL, LH, HH) capturing the horizontal, vertical, and diagonal details, respectively. The 2D Discrete Wavelet Transform (DWT) can be expressed mathematically as:

$$W_{j,k} = \iint f(x,y)\varphi_{j,k}(x,y)dxdy \quad (1)$$

Here $f(x,y)$ is the input image, $\varphi_{j,k}(x,y)$ is the wavelet basis function, and $W_{j,k}$ are the wavelet coefficients representing the features at specific scales and positions. Similarly, the original image can be reconstructed using the Inverse Discrete Wavelet Transform (IDWT):

$$f(x,y) = \sum_{j,k} W_{j,k}\varphi_{j,k}(x,y) \quad (2)$$

Among various wavelet bases, such as Haar and Daubechies wavelets, Haar wavelets were selected in this study due to their computational simplicity and efficiency. The transform is implemented by applying low-pass ($H_0$) and high-pass ($H_1$) filters along the rows and columns of the image, followed by 2×2 downsampling, as illustrated in **Fig. 3**.

To process the OCT retinal images, we implemented Haar DWT in PyTorch, decomposing the images into four subbands, as shown in **Fig. 4**. The LL subband, serving as the low-frequency component, represents the approximation of the image and includes essential structural details, such as cystic fluid regions with pathological significance. However, this component suppresses retinal edge details and lacks fine-grained information vital for accurate classification.

In contrast, the LH, HL, and HH subbands provide high-frequency components enriched with edge and texture details but are often mixed with noise. To form a comprehensive high-frequency representation, these three subbands are summed, ensuring that critical edge information is retained while minimizing redundancy.

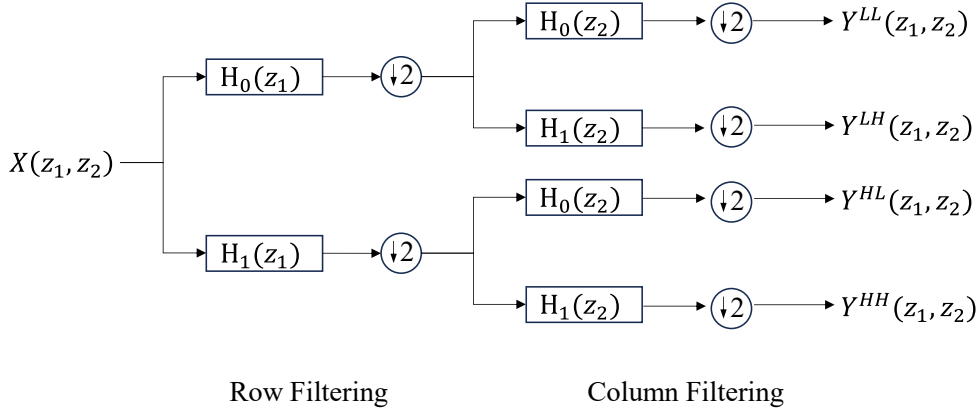

**Fig. 3** $H_0$ and $H_1$ denote the low-pass and high-pass filters, respectively. The input image is first filtered along the horizontal (row) direction using $H_0$ and $H_1$, producing two intermediate components. These components are then filtered along the vertical (column) direction to obtain four subbands (LL, LH, HL, and HH). After the wavelet transform, the image undergoes 2x downsampling.

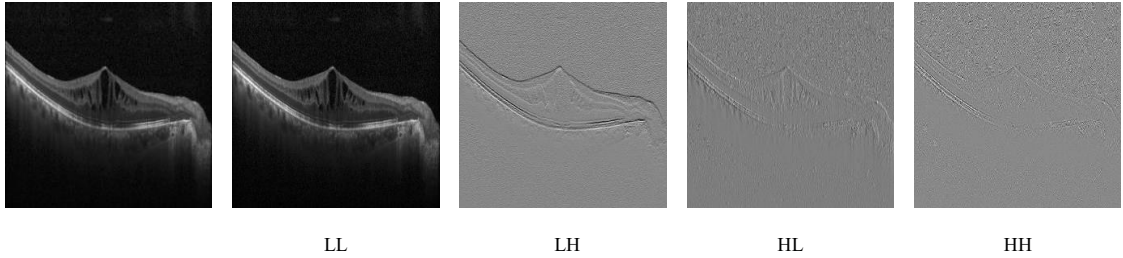

LL  LH  HL  HH

**Fig. 4** The result of the Haar Wavelet Transform on the OCT retinal image. For display purposes, we resized the decomposed images to match the original image size.

### 3.3. Multi-Scale Wavelet Spatial Attention (MSW-SA)

The receptive field is a key factor in spatial attention mechanisms. Previous studies have shown that an overly large receptive field fails to effectively capture small objects [40], while an overly small receptive field cannot fully utilize the contextual information within the image [29]. Additionally, some studies have shown that when dilated convolutions are used to expand the receptive field, the effect of spatial attention reaches saturation [30]. This may be due to two reasons: i) The large receptive field provided by the large kernel can lead to over-parameterization; ii) Large kernels often require a significant amount of padding to maintain the size of the feature map, and excessive padding can hinder the effectiveness of spatial attention.

Inspired by [38], we use wavelet convolutions with an efficient expansion of the receptive field as our tool, and we perceive spatial information using multi-scale receptive fields rather than a single scale. Given an input feature map $F \in R^{H \times W \times C}$, a spatial attention map $S(F) \in R^{H \times W \times 1}$ can be derived, and we replicate it along the channel dimension to obtain a 3D spatial attention map $S'(F) \in R^{H \times W \times C}$. After passing through the Multi-Scale Wavelet Spatial Attention module, the output feature map is:

$$F' = F + F \times S'(F) \qquad (3)$$

Here, $F'$ denotes the original feature map, and $S'(F)$ represents the corresponding feature map.

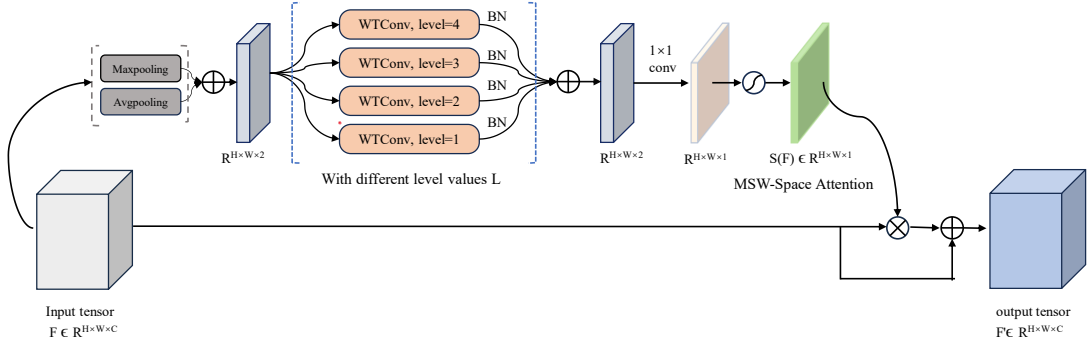

**Fig. 5** Implementation of MSW-SA when the feature map size is 28×28. Here, "level" refers to the number of wavelet decomposition stages. Wavelet convolution decomposes the feature map into different frequency sub-bands and performs convolution on the LL component obtained from each decomposition. From the second decomposition onward, only the LL component from the previous stage is further decomposed. This process is repeated until the specified number of levels is reached, after which the feature map is reconstructed from the decomposition results in a bottom-up manner.

Firstly, we employ average pooling and global pooling to compress the feature map, thereby integrating inter-channel information and reducing computational complexity. Subsequently, the compressed feature maps are concatenated to represent the information of the original feature map. Next, we use progressively increasing wavelet convolutions to achieve a global receptive field. The contextual information from different receptive fields is used to perceive the spatial locations of lesions of different sizes. As shown in **Fig. 5**, when the feature map size is 28×28, we apply wavelet convolutions up to level 4. After each wavelet convolution, a batch normalization (BN) layer is added to balance the contribution of each branch. The outputs of the different branches are then summed to aggregate the spatial information, which is subsequently passed through a 1×1 convolution to reduce the feature map's dimensionality to 1. Finally, a Sigmoid function is applied to obtain the spatial attention map S(F). In summary, the computation of our Multi-Scale Wavelet Spatial Attention is as follows:

$$S(F) = \text{Sigmoid}(W_{Cov}(\sum_{i=1}^{n} \text{BN}(W_T^n(\text{Max}(F) + \text{Avg}(F))))) \qquad (4)$$

where $W_{cov}$ denotes a 1×1 normal convolution, BN represents the batch normalization layer, Max and Avg indicate channel-wise max pooling and average pooling, respectively. While $W_T^n$ refers to the n-level wavelet convolution, The value of n depends on the size of the feature map.

### 3.4. High-frequency Feature Compensation Block (HFFC)

The HFFC block aims to filter out high-frequency features useful for classification from the high-frequency components containing significant noise after wavelet decomposition, using a frequency learning approach, and leverage these features to assist classification. Initially, convolutional layers are used to extract local spatial features while increasing the number of channels. After passing through activation functions and Batch Normalization (BN) layers, the feature maps undergo the most crucial process—High-Frequency Effective Feature Extraction (HFE). Finally, convolution is applied to integrate the feature maps, followed by max pooling to reduce their size. Since wavelet transformation may cause negative values in the feature maps initially, LeakyReLU is used as the activation function to prevent neuron death during the early stages of training. The core task of HFE is to distinguish between noise features and high-frequency features useful for classification in the frequency domain. This process mainly involves two key steps: (1) Frequency Dynamic Decomposition of feature maps, and (2) Local Selection and Global Fusion of feature maps.

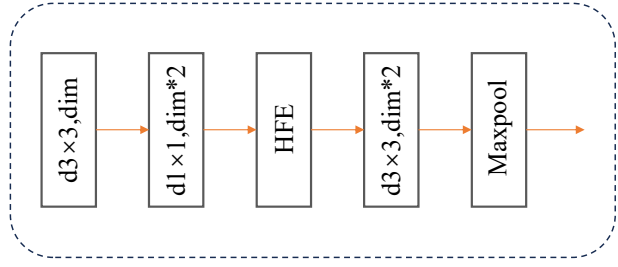

**Fig. 6** Architecture of the HFFC Block

Frequency Dynamic Decomposition of feature maps: Considering the complex differences between noise features and high-frequency features useful for classification, using frequency transformation tools with a fixed decomposition approach may not be optimal. We use a learnable low-pass filter to generate low-frequency feature maps. Specifically, the learnable low-pass filter can be obtained using the following Equation (5):

$$low_{filter} = \text{Softmax}(\text{BN}(\text{Conv}(\text{GAP}(x)))) \tag{5}$$

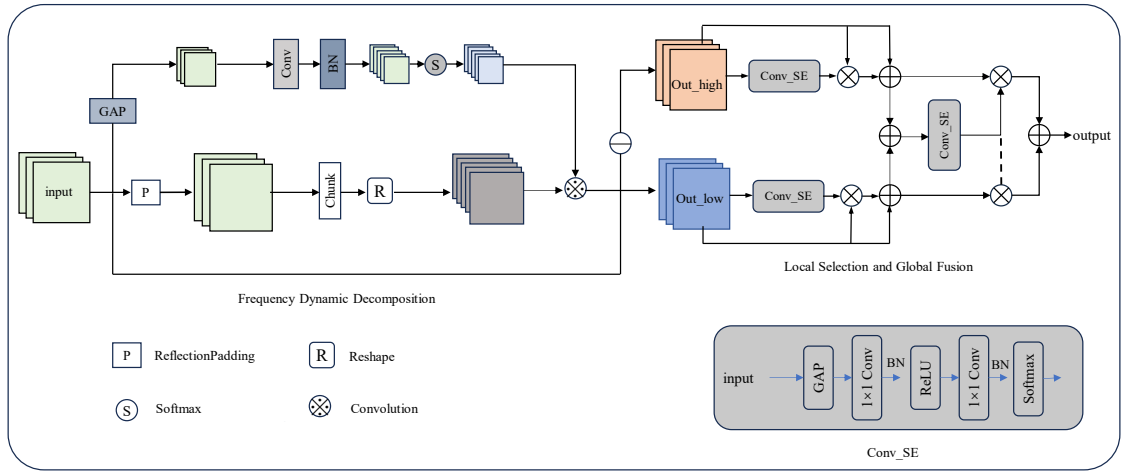

**Fig. 7** The process of FFE. The FFE process consists of two parts: dynamic decomposition of feature maps and frequency-selective fusion. First, the feature maps are decomposed into different frequency components using a designed dynamic filter. Next, the corresponding high-frequency and low-frequency components are locally weighted via a channel attention mechanism. The weighted feature maps are then fused to obtain global fusion information. Finally, the global fusion information is used to generate weights for the different frequency components, and the weighted feature maps are summed to produce the final output.

Firstly, we apply reflection padding to the feature maps to fully utilize the edge information. Then, the input feature map is split into multiple groups $x'$, and the filter is convolved with the grouped feature maps to obtain the low-frequency feature map (Equation (6)):

$$y_{out_{low}} = x' * low_{filter} \tag{6}$$
$$y_{out_{high}} = x - y_{low} \tag{7}$$

The high-frequency feature map is obtained from Equation (7), where $x$ is the input feature map, $x'$ is the padded and grouped feature map, $y_{out_{low}}$ is the low-frequency output, and $y_{out_{high}}$ is the high-frequency output. Here, * denotes convolution

Local Selection and Global Fusion of feature maps: We use a channel attention mechanism to weight the feature maps from different frequencies, emphasizing the features that are useful for classification, and perform global fusion through soft weighting. Overall, we follow the design of Squeeze-and-Excitation (SE) networks [41], replacing fully connected layers with pointwise convolutions.

We use a channel attention mechanism to weight the feature maps from different frequencies, emphasizing high-frequency features that are useful for classification. In general, we follow the design principles of Squeeze-and-Excitation (SE) networks, replacing fully connected layers with pointwise convolutions to implement the channel attention mechanism. We apply local selection weighting to both low-frequency and high-frequency feature maps, aggregating them separately to obtain local channel context information. This local context information is then processed through the channel attention module to obtain weight information, thereby retaining the most effective low-frequency and high-frequency features. To make feature selection in the frequency domain more precise and overcome potential semantic differences between low-frequency and high-frequency feature maps, we perform global fusion after the local selection. The global fusion uses a soft weighting approach. This design decision is based on the consideration that the high-frequency compensation network should not be over-parameterized, as excessive parameters could cause the network to overfit noise features and other high-frequency features, rather than selecting high-frequency features that are beneficial for classification. Specifically, the locally weighted low-frequency and high-frequency feature maps are globally fused through summation for channel context aggregation, and then passed through the channel attention module to obtain global fusion weights. The feature map resulting from global fusion can be expressed by Equation (8), (9):

$$C(F) = \text{Softmax}(BN(Conv(ReLU(BN(Conv(GAP(F)))))))\qquad(8)$$

$$Global_{fusion} = L_{Local} \times C(L_{Local} + H_{Local}) + H_{Local} \times [1 - C(L_{Local} + H_{Local})]\qquad(9)$$

where $L_{Local}$ and $H_{Local}$ represent the locally weighted low-frequency and high-frequency features, respectively, and C(F) is the function used to obtain the fusion weights. This approach allows the model to make a soft selection between high-frequency and low-frequency feature maps. Through local selection weighting and global fusion, the module can fully utilize the frequency differences between noise and valid high-frequency features.

## 4. Experiments and Results

### 4.1. Dataset

OCT-C8 [42]: The dataset contains 24,000 optical coherence tomography (OCT) retinal images, covering eight types of retinal diseases: AMD, CNV, CSR, DME, DR, Drusen, MH, and normal (3,000 images per class). The image sizes range from 384×496 pixels to 1536×496 pixels, with an average size of 750×500 pixels. The dataset is divided into three subsets: training, validation, and testing. The training set contains 18,400 images, while both the validation and testing sets contain 2,800 images each, with an equal number of images from the eight categories in each subset.

OCT2017 [43]: The dataset, provided by the University of California, San Diego, consists of 84,484 OCT retinal images. It is divided into four categories: CNV (37,455 images), DME (11,598 images), Drusen (8,866 images), and normal (26,565 images). The image sizes range from 433×289 pixels to 496×1536 pixels, with an average size of 498×682 pixels. All images are labeled with the lesion type, patient ID, and image number. The dataset is divided into training and test sets, with 83,484 images in the training set and 250 images per category in the testing set, totaling 1,000 images.

## 4.2. Evaluation metrics

To evaluate the effectiveness of the model's performance, we choose Accuracy (ACC), Precision, Sensitivity and F1-score as classification metrics, which can be calculated by the confusion matrix. We define: True Positive (TP), True Negative (TN), False Positive (FP), and False Negative (FN). The ACC, Precision, and Sensitivity can be calculated by the following equation.

$$Accuracy = \frac{TP + TN}{TP + TN + FP + FN} \tag{10}$$

$$Precision = \frac{TP}{TP + FP} \tag{11}$$

$$Sensitivity = \frac{TP}{TP + FN} \tag{12}$$

Precision reflects the accuracy of the model's predictions, while sensitivity reflects the comprehensiveness of the model's predictions. The F1 score is a metric used to balance precision and sensitivity, and it can be calculated by Equation (13). A higher value indicates better performance.

$$F1\ score = \frac{2 \times Precision \times Sensitivity}{Precision + Sensitivity} \tag{13}$$

## 4.3. Implementation details

The experiments were conducted on an NVIDIA RTX 3090 GPU. Before feeding the images into the model, all images were resized to 448×448 pixels and standardized to ensure stable training. To prevent overfitting and enhance the model's generalization ability, we employed basic data augmentation techniques available in PyTorch (see **Tab. 1**). We used the Adam optimizer with momentum parameters $\beta_1=0.9$ and $\beta_2=0.999$, and applied a cosine annealing learning rate scheduler with a warm-up phase. The base learning rate was set to 2e-4, and the minimum learning rate was set to 2e-6. Weight decay was set to 1e-4. For the OCT-C8 dataset, the batch size was set to 32, and the model was trained for 60 epochs. The model that performed best on the validation set was selected for testing, and 5 independent experiments were conducted. In each experiment, the model was trained 5 times under the same settings, and testing was performed 5 times. The average value and standard deviation of the 5 test results were then computed. For the OCT2017 dataset, a batch size of 32 was used, and 5-fold cross-validation was applied, with each fold trained for 30 epochs. The results from the 5 folds were averaged, and the standard deviation was calculated to obtain a more robust performance metric. The average value and standard deviation of the data can be calculated using the following equations:

$$\mu = \frac{1}{N} \sum_{i=1}^{N} X_i \tag{14}$$

$$\delta = \sqrt{\frac{1}{N} \sum_{i=1}^{N} (X - u)^2} \tag{15}$$

where N is the number of samples, is the value of the i-th sample, $\mu$ is the average value, and $\delta$ is the standard deviation.

Furthermore, we performed a paired t-test to assess the performance differences between our model and others. The p-value indicates the significance of the difference, with values below 0.05 considered statistically significant. Bold indicates the best performance, underline indicates p <0.05 (statistically significant). Our code is publicly available on https://github.com/liuliutaotao64/WaveNet-SF

**Tab. 1** Data Augmentation in our study

| Data Augmentation | values |
|:---:|:---:|
| RandomRotation | ±15° |
| RandomResizedCrop | (0.8,1.2) |
| ColorJitter | 20% |
| RandomHorizontalFlip | True |

**4.4. The Performance of Our Proposed Approach**

In this section, we compare the performance of our model with that of state-of-the-art general deep convolutional neural network methods, as well as other cutting-edge models specifically proposed for OCT retinal disease monitoring, on the OCT-C8 and OCT2017 datasets. FPN-ResNet50 and FPN-DenseNet121 were initially proposed by Sotoudeh-Paima, Jodeiri [19], who combined ResNet50[44] and DenseNet121[45] with a feature pyramid network (FPN), using concatenation of feature maps at different scales to enhance classification performance. ResNet-EdgeEn[17] increases the derivative contrast by adopting a novel cross-activation function, thereby focusing more on the retinal region rather than the background. Swim-Poly Transformer[16] optimizes the cross-entropy loss by adjusting the weights of the polynomial basis, effectively modeling multi-scale features. HTC-Retina[15] combines convolutional neural networks (CNN) and vision transformers (VIT) to capture local features and learn global dependencies. To further enhance the comparison, we also include VGG16 [46], GoogLeNet [47], InceptionV3 [48],and EfficientNet-B3 [49], all of which have demonstrated exceptional performance on various vision tasks. For the methods in Ref.[15-17], we directly adopted the results presented in their respective papers, as these works use the same data, the same data partitioning, and the same task, making a direct comparison valid. For the other methods, we have re-implemented them under our experimental conditions.

*4.4.1. Results on the OCT-C8 Dataset*

**Tab. 2** presents the performance comparison of our method with several state-of-the-art approaches on the OCT-C8 dataset. Our method achieved an accuracy of 97.82%, outperforming all the other models listed in the table. Among the compared models, FPN-DenseNet121 achieved an accuracy of 97.22%, followed by DenseNet121 with 97.21%, and FPN-ResNet50 with 97.14%. EfficientNet-B3 and Swim-Poly Transformer achieved accuracies of 97.12% and 97.11%, respectively, while InceptionV3 and GoogLeNet achieved 97.10% and 96.79%, respectively. Notably, ResNet-EdgeEn showed the lowest accuracy of 92.40%, with lower precision, sensitivity, and F1 score as well. In terms of precision, sensitivity, and F1 score, our method also outperformed all the others, achieving 97.83%, 97.82%, and 97.82%, respectively. This confirms the robustness of our approach across multiple evaluation metrics. The results highlight that our model excels in comparison to other state-of-the-art models in terms of both accuracy and other performance measures. In addition, the results showed that all *p*-values were less than 0.05, indicating that the performance improvement of our method is statistically significant.

Tab. 2 Performance comparison of the OCT-C8 dataset

| Method | Accuracy | Precision | Sensitivity | F1 score | P-value | Parameter | Inference-Time |
|---|---|---|---|---|---|---|---|
| VGG16[46] | 96.45±0.21 | 96.47±0.20 | 96.45±0.21 | 96.45±0.21 | 3.06×10$^{-4}$ | 134.29M | 0.3264s |
| ResNet50[44] | 96.98±0.13 | 97.00±0.13 | 96.98±0.13 | 96.98±0.13 | 1.36×10$^{-3}$ | 23.52M | 0.3357s |
| GoogLeNet[47] | 96.79±0.14 | 96.80±0.14 | 96.79±0.14 | 96.78±0.15 | 3.68×10$^{-4}$ | **6.62M** | 0.4375s |
| InceptionV3[48] | 97.10±0.12 | 97.11±0.15 | 97.10±0.15 | 97.10±0.15 | 1.91×10$^{-3}$ | 23.83M | 0.3747s |
| DenseNet121[45] | 97.21±0.13 | 97.22±0.13 | 97.21±0.13 | 97.21±0.13 | 5.49×10$^{-3}$ | 6.96M | 0.4368s |
| EfficientNet-B3[49] | 97.12±0.21 | 97.13±0.21 | 97.12±0.21 | 97.12±0.21 | 2.07×10$^{-3}$ | 12.23M | **0.3225s** |
| FPN-ResNet50[18] | 97.14±0.25 | 97.15±0.25 | 97.14±0.25 | 97.14±0.25 | 7.83×10$^{-3}$ | 27.38M | 0.3736s |
| FPN-DenseNet121[19] | 97.22±0.13 | 97.23±0.12 | 97.22±0.13 | 97.21±0.13 | 9.77×10$^{-3}$ | 9.81M | 0.4981s |
| ResNet-EdgeEn[17] | 92.40 | 93.00 | 92.00 | 92.00 | - | - | - |
| Swin-Poly-Transformer[16] | 97.11 | 97.13 | 97.11 | 97.10 | - | - | - |
| HTC-Retina[15] | 97.00 | 97.04 | 97.00 | 97.01 | - | - | - |
| **Ours** | **97.82±0.19** | **97.83±0.19** | **97.82±0.19** | **97.82±0.19** | - | 27.84M | 0.5118 |

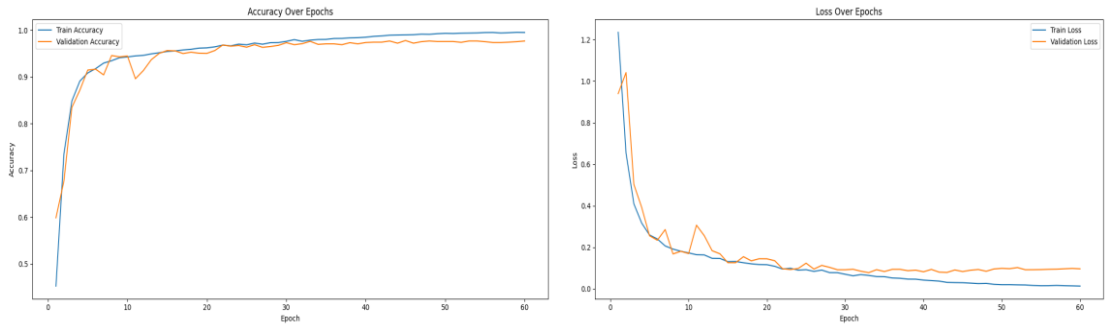

**Fig. 8** Loss and accuracy during the training process

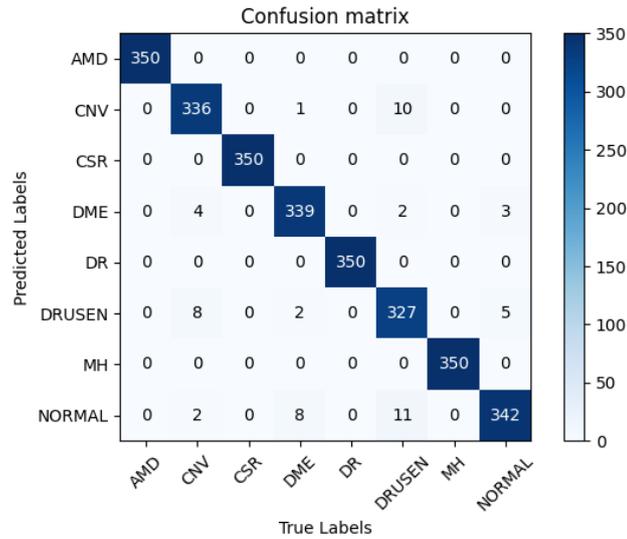

**Fig. 9** The confusion matrix of the results, with an accuracy of 98.00%

**Fig. 8** illustrates the training process of our method. On the training set, the loss continuously decreased while the accuracy gradually improved. On the validation set, the loss initially decreased and then stabilized, while the accuracy increased and eventually plateaued, with no signs of overfitting. **Fig. 9** presents the confusion matrix from one of the experiments, showing that all classes were accurately classified, particularly AMD, CSR, DR, and MH, which achieved 100% classification accuracy. The other classes in the confusion matrix also demonstrated high classification precision, further validating the effectiveness of our method.

*4.4.2. Results on the OCT2017 Dataset*

**Tab. 3** Performance comparison of the OCT2017 dataset

| Method | Accuracy | Precision | Sensitivity | F1 score | P-value |
|---|---|---|---|---|---|
| VGG16[46] | 99.52±0.17 | 99.53±0.16 | 99.52±0.17 | 99.52±0.17 | $2.50\times10^{-2}$ |
| ResNet50[44] | 99.26±0.12 | 99.28±0.11 | 99.26±0.12 | 99.26±0.11 | $\underline{5.13\times10^{-4}}$ |
| GoogLeNet[47] | 99.40±0.16 | 99.41±0.16 | 99.40±0.16 | 99.40±0.16 | $9.74\times10^{-2}$ |
| InceptionV3[48] | 99.28±0.24 | 99.30±0.23 | 99.28±0.24 | 99.28±0.24 | $5.95\times10^{-2}$ |
| DenseNet121[45] | 99.26±0.24 | 99.27±0.23 | 99.26±0.24 | 99.26±0.24 | $8.81\times10^{-2}$ |
| EfficientNet-B3[49] | 99.12±0.11 | 99.14±0.10 | 99.12±0.11 | 99.12±0.11 | $\underline{1.77\times10^{-3}}$ |
| FPN-ResNet50[18] | 99.34±0.10 | 99.35±0.09 | 99.34±0.10 | 99.34±0.10 | $\underline{3.04\times10^{-2}}$ |
| FPN-DenseNet121[19] | 99.32±0.20 | 99.33±0.19 | 99.32±0.20 | 99.32±0.20 | $8.63\times10^{-2}$ |
| HTC-Retina[15] | 99.40 | 99.39 | 99.41 | 99.40 | |
| **Ours** | **99.58±0.13** | **99.58±0.12** | **99.58±0.13** | **99.58±0.13** | |

In order to further evaluate the generalizability of our model across diverse retinal disease datasets, we conducted an extensive evaluation on the OCT2017 dataset. **Fig. 10** presents the loss and accuracy curves from the 5-fold cross-validation. As shown, the model exhibited a consistent decrease in loss and a gradual increase in accuracy across all folds during training, demonstrating the effectiveness of our approach in learning discriminative features. On the validation set, the loss initially decreased and stabilized, while the accuracy continued to rise and eventually plateaued, with no indications of overfitting, further affirming the model's robustness and stability. **Fig. 11** illustrates the confusion matrix from the 5-fold cross-validation. The model achieved perfect classification for the CNV class, and other classes also demonstrated high classification accuracy. Specifically, the accuracies for each fold were as follows: Fold 1: 99.80%, Fold 2: 99.40%, Fold 3: 99.50%, Fold 4: 99.60%, and Fold 5: 99.60%, showcasing the model's consistent performance across different data splits. **Tab. 3** compares our method to other state-of-the-art models on the OCT2017 dataset. Among all methods, VGG16 achieved the second highest accuracy of 99.52%, while our method still outperformed it, achieving an accuracy of 99.58%. Additionally, our method also surpassed VGG16 in terms of precision, sensitivity, and F1 score, further underscoring its superior performance. As shown in the results, the *p*-values for some models in the OCT2017 dataset, such as GoogLeNet, InceptionV3, DenseNet121, and FPN-DenseNet121, exceed 0.05. This may be attributed to the relatively simple four-class classification task in OCT2017, which is less challenging than the complex eight-class task in OCT-C8. Additionally, the large training set size of OCT2017 (83,484 images) further limits the potential for performance improvement.

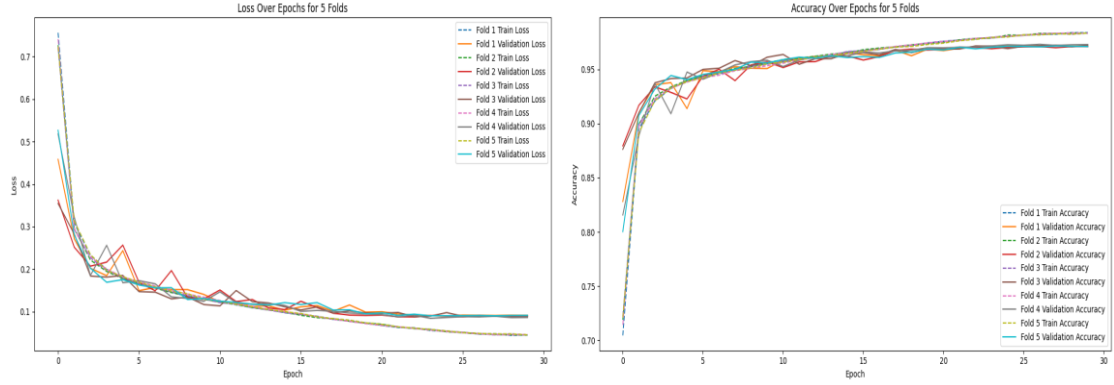

**Fig. 10** Loss and accuracy during the training process

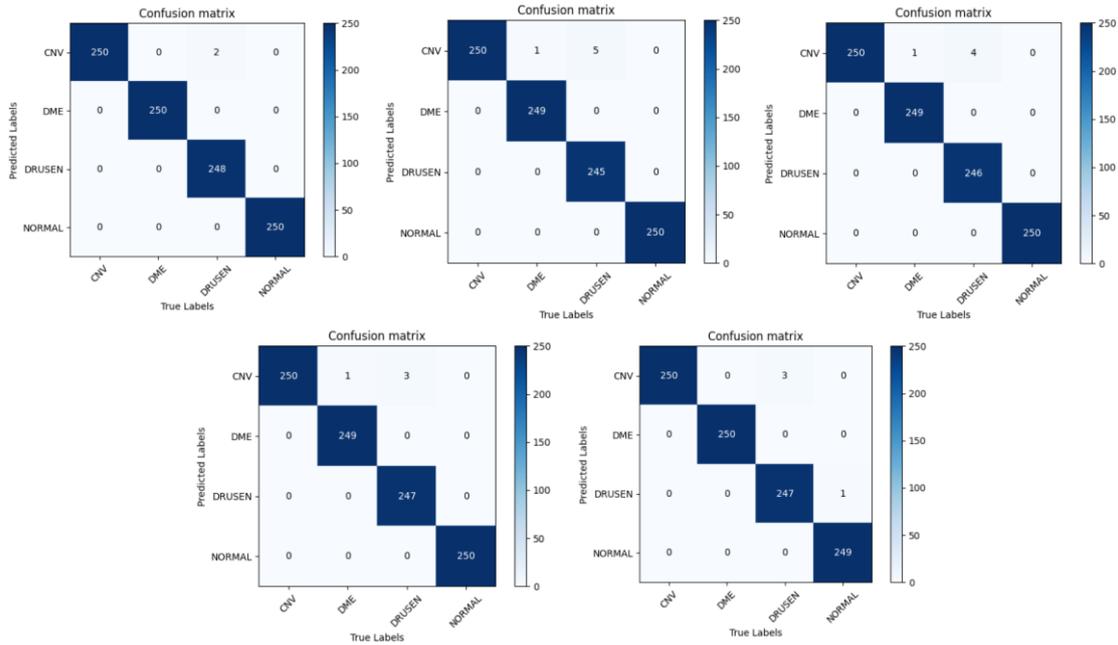

**Fig. 11** The confusion matrix for 5-fold cross-validation with accuracies as follows: Fold 1: 99.8%, Fold 2: 99.4%, Fold 3: 99.5%, Fold 4: 99.6%, Fold 5: 99.6%.

**4.5. Ablation Study**

To evaluate the individual contributions of each component in our proposed model, we conducted ablation experiments on the OCT-C8 dataset. As shown in **Tab. 4**, we used ResNet50 as our baseline model and progressively integrated different components to form the complete WaveNet-SF network. Initially, we incorporated the low-frequency components obtained through wavelet transform, which are denoised and retain important global structural information. This modification led to an accuracy of 97.42%, marking a 0.44% improvement over the baseline. Next, we added the Multi-Scale Wavelet Spatial Attention (MSW-SA) module, which was inserted between each stage of ResNet50. This addition resulted in a further improvement of 0.19%, achieving an accuracy of 97.61%. Finally, we introduced the High-Frequency Feature Compensation (HFFC) block to recover the high-frequency features that are typically lost during wavelet decomposition. This final modification (WaveNet-SF) led to an additional 0.21% increase in accuracy, culminating in a total accuracy of 97.82%, the highest among all configurations.

Tab. 4 Ablation experiment results on OCT-C8 dataset

| Method | Accuracy | Precision | Sensitivity | F1 | Parameter | Inference-Time |
|---|---|---|---|---|---|---|
| ResNet50 | 96.98±0.13 | 97.00±0.13 | 96.98±0.13 | 96.98±0.13 | **23.52M** | **0.3357s** |
| ResNet50+Wavelet | 97.42±0.11 | 97.44±0.10 | 97.42±0.11 | 97.41±0.11 | **23.52M** | - |
| ResNet50+Wavelet+MSW-SA | 97.61±0.15 | 97.62±0.15 | 97.61±0.15 | 97.61±0.15 | 23.53M | - |
| **WaveNet-SF (Ours)** | **97.82±0.19** | **97.83±0.19** | **97.82±0.19** | **97.82±0.19** | 27.84M | 0.5118s |

To further validate the effectiveness of the proposed Multi-Scale Wavelet Spatial Attention (MSW-SA), we conducted a comparative experiment by replacing MSW-SA with two well-established attention mechanisms: BAM [30] and CBAM [29], both of which are hybrid attention mechanisms that combine spatial and channel attention. Since our MSW-SA is purely a spatial attention mechanism, we only compared our method with the spatial attention components of BAM and CBAM, effectively removing their channel attention parts for this experiment. In **Tab. 5**, we present a comparison of the three attention mechanisms in terms of classification accuracy and the total number of parameters.

Tab. 5 Comparative Experiment of BAM, CBAM, and MSW-SA on the OCT-C8 Dataset

| Method | Accuracy | Precision | Sensitivity | F1 | Parameters |
|---|---|---|---|---|---|
| BAM[30] | 97.69±0.17 | 97.70±0.17 | 97.69±0.17 | 97.69±0.17 | 28.02M |
| CBAM[29] | 97.51±0.19 | 97.52±0.19 | 97.51±0.19 | 97.51±0.19 | 27.84M |
| **MSW-SA (Ours)** | **97.82±0.19** | **97.83±0.19** | **97.82±0.19** | **97.82±0.19** | **27.84M** |

From **Tab. 5**, it is evident that BAM achieved better performance than CBAM, but it comes at the cost of a higher parameter count. On the other hand, our MSW-SA, with a parameter count comparable to CBAM, outperforms both BAM and CBAM in terms of classification accuracy. This suggests that our MSW-SA not only enhances performance but also maintains a lightweight architecture, making it a more efficient choice for retinal disease classification.

In addition, we conducted supplementary experiments to evaluate the impact of each component within the High-Frequency Feature Compensation (HFFC) module and the importance of frequency-domain learning. The HFFC module consists of two key steps: frequency dynamic decomposition of feature maps and local selection and global fusion of feature maps, which are the key distinctions from spatial-domain learning. In Experiment 1 (HFFC without Local Selection and Global Fusion), we removed the local selection and global fusion process, retaining only the frequency dynamic decomposition of feature maps. Specifically, the feature maps were decomposed into multiple frequency components, which were then directly summed as output. In Experiment 2 (HFFC without Frequency Dynamic Decomposition), we performed the opposite operation by removing the frequency dynamic decomposition and directly applying the local selection and global fusion to the input feature maps, focusing on pure spatial-domain feature learning. In Experiment 3 (HFFC without FFE), we removed the entire FFE submodule in the HFFC module, namely both the frequency dynamic decomposition and the local selection/global fusion processes, resulting in the HFFC degrading into a purely convolutional spatial-domain network. Experiment 4 (HFFC without FFE but deeper) was based on Experiment 3, with an appropriate increase in the number of convolutional layers, in order to evaluate whether a deeper spatial convolutional structure alone, without introducing frequency-domain mechanisms, could achieve comparable performance.

Tab. 6 Ablation Experiment of HFFC

| Method | Accuracy | Precision | Sensitivity | F1 | Parameters |
| --- | --- | --- | --- | --- | --- |
| HFFC without Local Selection and Global Fusion | 97.33±0.09 | 97.37±0.10 | 97.33±0.07 | 97.34±0.09 | 27.84M |
| HFFC without Frequency Dynamic Decomposition | 97.35±0.10 | 97.39±0.10 | 97.35±0.10 | 97.35±0.10 | **27.63M** |
| HFFC without FFE | 97.38±0.09 | 97.41±0.09 | 97.38±0.09 | 97.38±0.09 | **27.63M** |
| HFFC without FFE but deeper | 97.30±0.15 | 97.33±0.15 | 97.30±0.15 | 97.29±0.15 | 29.20M |
| Ours | **97.82±0.19** | **97.83±0.19** | **97.82±0.18** | **97.82±0.19** | 27.84M |

As shown in **Tab. 6**, when only the frequency dynamic decomposition of feature maps is performed, the accuracy is 97.33±0.09; when only local selection and global fusion are applied, the accuracy is 97.35±0.10. Both are lower than the accuracy of 97.82±0.19 achieved when both processes are combined. This indicates that when frequency dynamic decomposition or local selection and global fusion are used alone, the HFFC module not only fails to enhance detection capability by supplementing edge information but may also introduce interfering noise, leading to a degradation in model performance.

Furthermore, as shown in Experiments 4 and 5, after removing the FFE—an essential frequency-domain learning mechanism—even increasing the number of convolutional filters to make the parameter count exceed that of our model, the performance is still lower than that of using only the low-frequency feature extraction network (97.61±0.15). This further confirms that directly using high-frequency components may introduce noise and impair the model's performance.

**4.6. Robustness of Our Approach to Noise**

OCT retinal images are generated by an optical coherence tomography (OCT) system. Due to interference signals caused by the backscattered light of biological tissues, the original OCT images often exhibit significant speckle noise. This noise weakens the edges and details of the image, making feature extraction more challenging for models. To evaluate the robustness of our model to speckle noise, we first need to establish a model that accurately simulates speckle noise in OCT retinal images. We adopted the widely accepted multiplicative speckle noise model, as described in [25], which can be represented by the following equation:

$$F(x,y) = g(x,y) + g(x,y) \times u(x,y) \tag{14}$$

where $g(x,y)$ represents the original image, $u(x,y)$ is Gaussian noise with a mean of 0 and variance $s$, which is related to the grayscale values of the original image: the higher the grayscale value, the higher the variance of the noise. $F(x,y)$ denotes the noisy image.

We introduced speckle noise with varying intensities into the OCT-C8 test dataset and OCT2017 test dataset. The noise intensity is determined by the Peak Signal-to-Noise Ratio (PSNR), with lower PSNR indicating higher noise intensity. The PSNR is calculated as follows:

$$\text{PSNR} = 20 \times \log_{10}\left(\frac{Max}{\sqrt{MSE}}\right) \tag{15}$$

where $Max$ is the maximum pixel value of the image, and MSE is the mean squared error between the noisy and original images. The model, trained on the noise-free training dataset, was then used to predict the noisy test dataset, and accuracy was reported to evaluate the robustness of our model and other methods under noise environment.

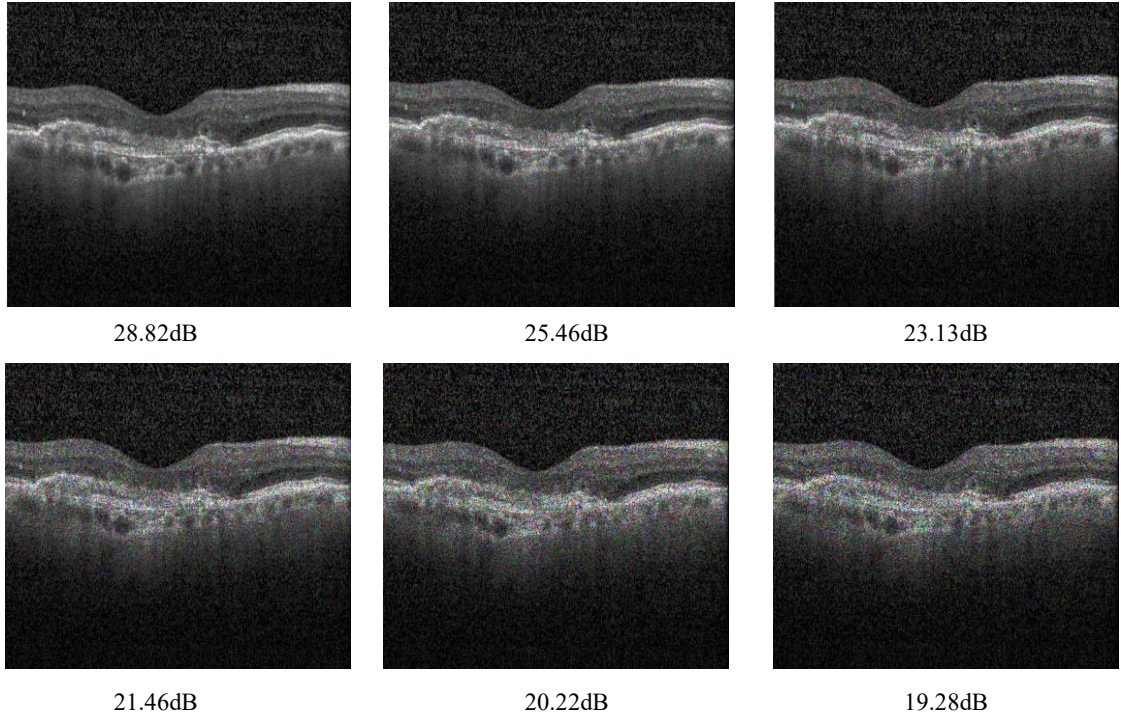

<p style="text-align:center">28.82dB     25.46dB     23.13dB</p>
<p style="text-align:center">21.46dB     20.22dB     19.28dB</p>

**Fig. 12** Visualization of image samples after adding speckle noise of varying intensities to the test dataset

**Fig. 12** shows the results of adding different levels of speckle noise to the same image in the test set. As the noise intensity increases, the image quality gradually deteriorates, and the PSNR (Peak Signal-to-Noise Ratio) value decreases, indicating that the level of image distortion intensifies. Specifically, when the PSNR value is low, it means that the noise interference becomes more pronounced, causing significant loss of detail and a substantial decrease in image quality.

The results in **Tab. 7** and **Tab. 8** highlight the superior noise robustness of our model across both the OCT-C8 and OCT2017 datasets. On the OCT-C8 dataset, as noise intensity increases (with decreasing PSNR values), our method consistently outperforms all other models. At the highest noise level (19.28dB), our model achieves an accuracy of 92.15%, which is only a 5.67% decrease from the baseline accuracy under noise-free conditions. This is significantly better than the performance of other methods, such as FPN-ResNet50, which experiences a 35.22% decrease, and DenseNet121, which sees a 35.14% decrease. The poorest results are observed in InceptionV3 and FPN-DenseNet121, with accuracies dropping to 59.41% and 61.78%, respectively, at this noise level, demonstrating the vulnerability of these models to speckle noise. In contrast, our approach demonstrates exceptional robustness, maintaining a high level of accuracy even under severe noise interference.

In the OCT2017 dataset, our model similarly demonstrates strong noise resilience. At the lowest PSNR (19.28dB), it achieves 94.78% accuracy, which corresponds to a modest 4.72% decrease from the noise-free accuracy. This is a significant improvement over other methods: FPN-DenseNet121 shows a 13.32% decrease, and InceptionV3 experiences a 10.78% decrease. These results highlight the superior noise robustness of our method, as it consistently maintains high classification accuracy even under challenging noisy conditions.

Tab. 7 Comparison of Noise Robustness Across Models on OCT-C8 test dataset

| Noise intensity / Method | 28.82dB | 25.46dB | 23.13dB | 21.46dB | 20.22dB | 19.28dB |
|---|---|---|---|---|---|---|
| VGG16[46] | 96.25±0.20 | 95.83±0.47 | 95.11±0.67 | 93.27±0.97 | 91.02±1.30 | 87.38±2.30 |
| ResNet50[44] | 96.71±0.24 | 96.28±0.18 | 95.05±0.16 | 92.22±0.77 | 87.46±1.45 | 80.31±2.02 |
| GoogLeNet[47] | 96.75±0.11 | 96.37±0.19 | 95.55±0.31 | 93.23±0.92 | 89.83±1.82 | 85.30±2.69 |
| InceptionV3[48] | 96.62±0.11 | 95.20±0.98 | 91.00±3.74 | 81.92±8.57 | 70.76±9.60 | 59.41±7.98 |
| DenseNet121[45] | 96.87±0.03 | 95.64±0.64 | 90.75±4.72 | 80.77±7.46 | 69.76±6.02 | 62.07±4.03 |
| EfficientNet-B3[49] | 96.87±0.23 | 96.12±0.41 | 94.28±0.97 | 89.50±3.47 | 81.20±8.06 | 70.12±8.34 |
| FPN-ResNet50[18] | 96.96±0.20 | 96.14±0.20 | 92.29±1.07 | 81.70±2.25 | 69.79±1.88 | 61.92±1.26 |
| FPN-DenseNet121[19] | 97.09±0.15 | 96.39±0.20 | 93.66±0.50 | 86.27±3.49 | 73.97±5.66 | 61.78±5.10 |
| **Ours** | **97.82±0.21** | **97.54±0.18** | **97.13±0.20** | **96.41±0.36** | **95.05±0.74** | **92.15±0.81** |

Tab. 8 Comparison of Noise Robustness Across Models on OCT2017 test dataset

| Noise intensity / Method | 28.82dB | 25.46dB | 23.13dB | 21.46dB | 20.22dB | 19.28dB |
|---|---|---|---|---|---|---|
| VGG16[46] | 99.42±0.09 | **99.38±0.22** | 98.88±0.24 | **98.20±0.39** | **96.66±0.71** | 93.82±0.77 |
| ResNet50[44] | 99.16±0.27 | 98.74±0.17 | 98.00±0.39 | 96.88±0.73 | 95.62±1.35 | 93.58±2.17 |
| GoogLeNet[47] | 99.40±0.11 | 99.06±0.19 | 98.54±0.35 | 97.52±0.46 | 95.38±1.02 | 93.48±1.26 |
| InceptionV3[48] | 99.06±0.33 | 98.54±0.47 | 97.56±0.44 | 95.50±0.96 | 92.18±1.50 | 88.50±2.27 |
| DenseNet121[45] | 99.10±0.24 | 98.66±0.40 | 96.88±1.10 | 94.68±1.75 | 90.88±3.03 | 86.54±4.29 |
| EfficientNet-B3[49] | 99.02±0.09 | 98.66±0.16 | 98.10±0.43 | 96.48±0.93 | 94.14±1.68 | 90.24±3.03 |
| FPN-ResNet50[18] | 99.22±0.27 | 98.52±0.54 | 97.84±1.02 | 96.62±1.31 | 94.20±2.13 | 91.32±3.12 |
| FPN-DenseNet121[19] | 99.10±0.31 | 98.74±0.43 | 97.18±0.46 | 94.74±0.95 | 90.82±1.61 | 86.00±2.64 |
| **Ours** | **99.50±0.17** | 99.28±0.18 | **98.92±0.27** | 97.90±0.31 | 96.26±0.48 | **94.78±1.43** |

### 4.8 Evaluation of Model Generalization Ability

To further evaluate the generalization ability of the proposed model, we constructed an independent external dataset, referred to as OCT-Ext8. OCT-Ext8 was assembled by sampling images from multiple publicly available OCT retinal image datasets, including OCTID[50] (containing AMD, CSR, DR, MH, and Normal), NEH[7] (containing AMD, DME, and Normal), Srinivasan2014[51] (containing AMD, DME, and Normal), and OCT2017[43] (containing CNV, Drusen, DME, and Normal). The OCTID, NEH, and Srinivasan datasets are significantly smaller in scale compared to OCT2017. To ensure that the OCT-Ext8 dataset includes images from diverse clinical sources, six categories (excluding CNV and Drusen) were randomly selected from these three datasets, while the CNV and Drusen classes were sourced from the larger OCT2017 dataset. We extracted 1,000 images for each category whenever possible; if the available number was insufficient, all images were retained. As a result, the OCT-Ext8 dataset consists of eight categories with the following image counts: CNV (1000), Drusen (1000), DME (1000), Normal (1000), AMD (1000), MH (102), DR (107), and CSR (102), resulting in a total of 5,311 images. The model trained on the OCT-C8 dataset was directly applied to OCT-Ext8 without any fine-tuning, and its classification accuracy was compared with that of baseline and other competing methods.

**Tab. 9 Classification performance on OCT-Ext8 for generalization testing**

| Method | Accuracy | Precision | Sensitivity | F1 | P-value |
|---|---|---|---|---|---|
| VGG16[46] | 71.02±0.19 | 76.52±3.15 | 71.02±0.14 | 64.45±0.14 | $7.23 \times 10^{-3}$ |
| ResNet50[44] | 70.04±0.54 | 77.90±0.47 | 70.04±0.54 | 64.13±0.80 | $4.56 \times 10^{-4}$ |
| GoogLeNet[47] | **72.40±0.51** | 73.03±0.10 | **72.40±0.51** | **65.95±0.57** | $9.36 \times 10^{-1}$ |
| InceptionV3[48] | 68.20±1.33 | 70.12±7.10 | 68.20±1.33 | 61.56±1.36 | $4.83 \times 10^{-3}$ |
| DenseNet121[45] | 69.60±1.26 | 76.48±1.55 | 69.60±1.26 | 62.99±1.23 | $1.47 \times 10^{-2}$ |
| EfficientNet-B3[49] | 70.08±3.49 | 75.52±4.37 | 70.08±0.34 | 63.63±0.25 | $1.00 \times 10^{-5}$ |
| FPN-ResNet50[18] | 70.05±2.04 | 77.93±0.84 | 70.05±0.24 | 63.53±1.96 | $8.64 \times 10^{-2}$ |
| FPN-DenseNet121[19] | 71.31.±0.34 | 76.78±1.10 | 71.31±0.34 | 64.69±0.36 | $8.81 \times 10^{-2}$ |
| **Ours** | 71.84±0.47 | **79.70±0.40** | 71.84±0.47 | 65.06±0.51 | |

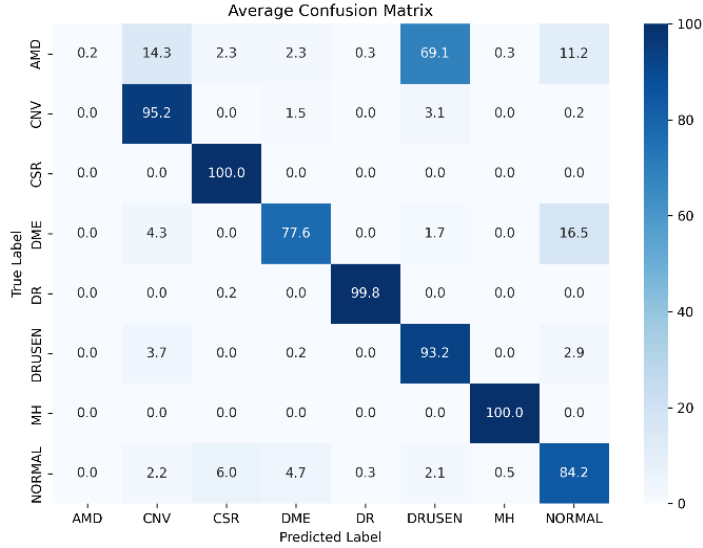

**Fig. 13** Confusion matrix averaged over five experimental runs

  **Tab. 9** presents the performance of all methods on the OCT-Ext8 dataset. As the proposed method is based on supervised learning, a decrease in accuracy is inevitable when it is directly applied in an unsupervised setting. Nevertheless, our method consistently outperforms all other competing approaches across all evaluation metrics, except for GoogLeNet. Compared with GoogLeNet, although our method exhibits slightly lower accuracy, sensitivity, and F1-score, it achieves significantly higher precision and demonstrates superior overall performance. These results further validate the strong generalization capability of the proposed method under diverse data conditions. Moreover, the *p*-values indicate that our model has statistically significant advantages over most other models. At the same time, although GoogLeNet surpasses our model in accuracy, it does not demonstrate statistical superiority.

  In **Fig. 13**, we present the confusion matrix of average per-class accuracy over five evaluation runs. Notably, the classification accuracy for the AMD class is considerably lower, likely due to a pronounced domain shift between the AMD samples in OCT-Ext8 and those in OCT-C8. Given that our approach is grounded in supervised learning, such performance variations are expected when confronted with domain shifts in a domain generalization setting.

## 5. Discussion

## 5.1 The Performance of WaveNet-SF

In recent years, multi-scale feature extraction techniques have gained prominence in the detection of retinal diseases from OCT images. These methods aim to enhance the model's capability to capture spatial information across various scales, which is crucial for accurate lesion detection. However, despite the advantages of multi-scale feature extraction, such approaches often face limitations in accurately identifying and localizing lesions. One major challenge is the lack of effective guidance from attention mechanisms, which results in imprecise focus on lesion regions, especially in complex retinal images where lesion sizes and shapes can vary considerably. Furthermore, many traditional methods fail to fully utilize retinal edge information, which is essential for distinguishing between healthy and affected regions. Additionally, OCT images are commonly affected by speckle noise, which is a type of interference caused by the scattering of light from biological tissues. This noise can obscure important features, making it difficult for existing methods to effectively detect and segment lesions.

To overcome these challenges, we propose a novel approach that leverages wavelet transform to decompose OCT images into low-frequency and high-frequency components. The low-frequency components preserve global structural information, which is critical for recognizing larger lesions and overall retinal anatomy, while simultaneously mitigating the impact of speckle noise. On the other hand, high-frequency components capture finer details, such as retinal edges, which are often vital for detecting small or subtle lesions. Guided by our Multi-Scale Wavelet Spatial Attention (MSW-SA), our model focuses more precisely on lesions of various sizes, shapes, and locations without relying on traditional multi-scale feature extraction methods. This targeted focus improves lesion localization accuracy and ensures that the model attends to the most clinically relevant features in OCT images. Furthermore, the High-Frequency Feature Compensation Network (HFFC) enhances the detection of key features, such as retinal edges, by compensating for information lost in the high-frequency domain, thus bolstering the model's performance in lesion detection.

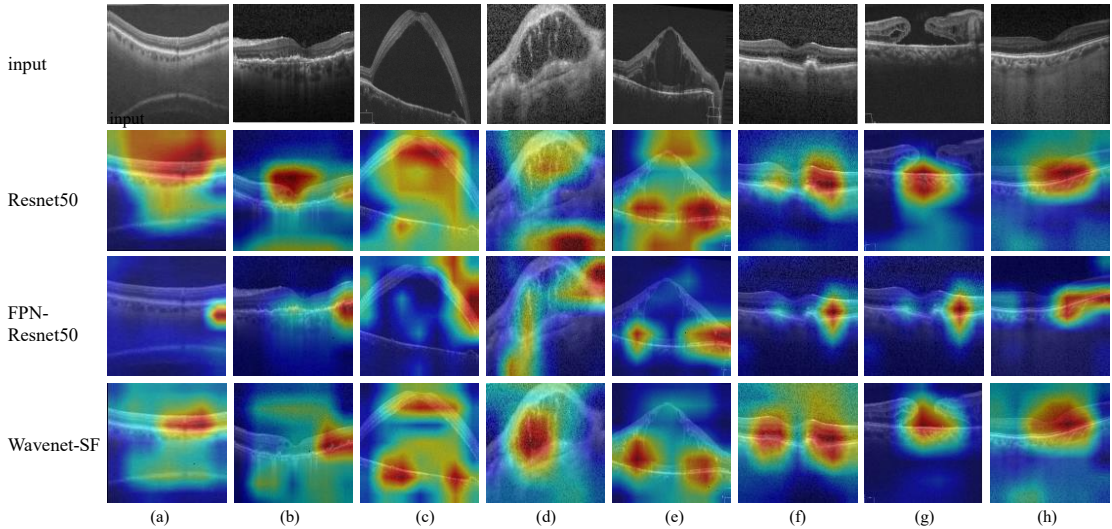

**Fig. 14** Heatmaps of retinal OCT images. (a) AMD (b) CNV (c) CSR (d) DME (e) DR (f) DRUSEN (g) MH (h) NORMAL

To provide an intuitive and interpretable evaluation of our model, we utilized the Grad-CAM method, which generates heatmaps reflecting the regions the model focuses on during prediction. By comparing the heatmaps generated by ResNet50 [44], the widely-used FPN-ResNet50 [18], and our WaveNet-SF model, we observed significant differences in performance. As illustrated in **Fig. 14**,

ResNet50 exhibits a tendency to focus on non-relevant regions, such as areas outside the retina, which compromises its ability to accurately identify lesions. In contrast, FPN-ResNet50 shows some improvement, with a more focused attention on lesion regions, but it still fails to capture the full extent of retinal lesions. Our model, on the other hand, demonstrates the highest precision in lesion localization, as evidenced by its ability to accurately identify the lower edge of the retina in **Fig. 14a**, even in challenging cases where other models struggle. Moreover, in cases like **Fig. 14c**, where significant morphological changes occur in the retina, our model successfully focuses on multiple lesion areas, showcasing its ability to handle complex retinal abnormalities. Notably, in **Fig. 14d**, which involves OCT images with severe speckle noise, both ResNet50 and FPN-ResNet50 struggle to effectively focus on the retina, while our model remains robust and accurately identifies the lesion regions. This further underscores the strength of our method in handling diverse lesion sizes, shapes, and noise levels, making it a more reliable choice for OCT retinal disease detection.

It is worth noting that although our method introduces a slight increase in both parameter size and inference time compared to the baseline model and the widely adopted FPN architecture for this task, these increments remain within an acceptable range for practical applications. Specifically, our model contains 27.84M parameters, which is only 4.32M more than the ResNet50 backbone (23.52M parameters), with an additional 0.1561s in inference time. When compared with FPN-ResNet50, which also employs ResNet50 as the backbone, our model incurs only 0.46M more parameters and an additional 0.1382s in inference time. More importantly, considering the substantial improvements achieved in baseline performance (Tables **2** and **3**), robustness to speckle noise (Tables **7** and **8**), and cross-domain generalization ability (Table **9**), these moderate increases are both reasonable and well justified.

**5.2 Ablation Study**

In Section 4.5, we demonstrated the effectiveness of each proposed method through ablation experiments. To gain a deeper understanding of the contributions of the two key components—MSW-SA and HFFC—we now turn to visualization techniques that offer a more intuitive analysis of their roles, and compare them with existing methods.

*5.2.1 Differences between MSW-SA, BAM, and CBAM*

The Multi-Scale Wavelet Spatial Attention (MSW-SA) mechanism enhances the model's detection ability by guiding it to focus on complex lesions within the low-frequency components of OCT retinal images, as shown in **Tab. 4**. Through extensive experimentation, we have confirmed that MSW-SA outperforms both BAM and CBAM for our specific task (**Tab. 5**). To provide a more intuitive understanding of the differences, we visualize the attention maps generated by each mechanism in the early layers of the model using heatmaps. These visualizations help clarify how MSW-SA differs from BAM and CBAM in its ability to focus on lesion regions.

As depicted in **Fig. 15**, in the early layers, CBAM tends to focus on a small, localized region of the image. This limited focus can be attributed to the use of a 7×7 convolutional kernel for spatial attention, which results in a small receptive field that restricts the model's ability to capture broader spatial information. In contrast, BAM uses two concatenated dilated convolutions with a kernel size of 3 and a dilation rate of 4, effectively expanding the receptive field and allowing for the capture of larger spatial contexts. However, this expansion introduces excessive padding, leading to the inclusion of irrelevant information, which ultimately compromises the accuracy of the attention mechanism. Consequently, while BAM covers a larger area in the early layers, its focus is less precise.

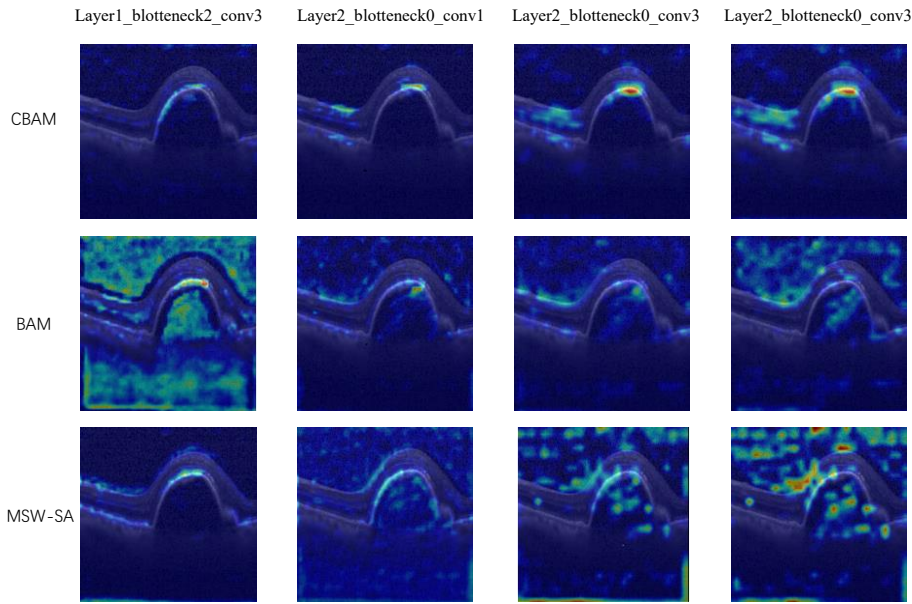

**Fig. 15** Visualization Analysis of CBAM, BAM, and MSW-SA

Our proposed MSW-SA, on the other hand, employs multi-scale receptive fields for spatial attention perception, gradually expanding to a global receptive field without resorting to excessive padding. This strategy enables MSW-SA to capture lesions of varying sizes with precision while maintaining the necessary global context for accurate lesion localization. As a result, MSW-SA not only focuses on a broader area but also achieves higher accuracy in focusing on lesions of different sizes and positions, as evidenced by the results discussed in Section 5.1. This efficient expansion of the receptive field, which avoids introducing excessive irrelevant information, not only improves the parameter efficiency of our design but also achieves an effective balance between local and global attention. Consequently, our model outperforms both BAM and CBAM in lesion detection and localization tasks.

*5.2.2 The function of HFFC*

The High-Frequency Feature Compensation Network (HFFC) plays a pivotal role in our model by selectively extracting and enhancing high-frequency details, such as retinal edges, from the high-frequency components of OCT retinal images. By learning in the frequency domain, HFFC improves the model's ability to detect subtle features that are crucial for accurate retinal disease classification (**Tab. 4**). HFFC consists of two core strategies: dynamic decomposition of feature maps and local selection coupled with global fusion. These strategies work together to refine image features that are vital for disease detection.

To elucidate the functionality of the High-Frequency Feature Compensation (HFFC) module, we conducted a detailed visual analysis of the feature maps processed by the network, highlighting the transformations induced by HFFC. These visualizations serve to illustrate how HFFC enhances critical image features and to reveal the detrimental effects on detection performance when either of its two core strategies—dynamic decomposition or local selection with global fusion—is applied in isolation.

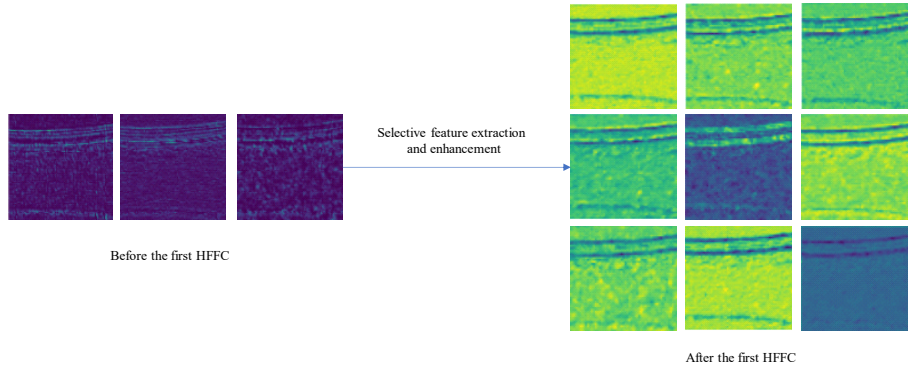

**Fig. 16** Visualization of feature maps of the first HFFC.

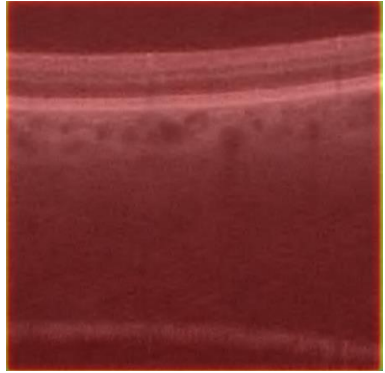

**Fig. 17** Heatmaps of the first HFFC.

**Fig. 16** illustrates the processing of feature maps by the first HFFC module. Prior to HFFC processing, retinal edge information is barely discernible, and substantial noise interferes with the feature representation. In contrast, following the first HFFC module, edge information is effectively enhanced while noise is suppressed. This improvement is achieved by decomposing the feature maps into distinct frequency components in the frequency domain, which facilitates the separation of edge details from noise. Subsequently, a lightweight local selection and global fusion strategy is applied, further reinforcing the edge features and mitigating noise. This design ensures that only relevant edge information is emphasized, thereby improving the model's capacity to focus on critical details.

**Fig. 17** presents the HFFC heatmap. Notably, the module appears to attend broadly to the overall image region. This observation reflects the inherent difference between frequency-domain learning and conventional spatial attention mechanisms: HFFC operates by decomposing the image into multiple frequency components, rather than emphasizing specific spatial regions. As a result, the heatmap does not directly convey the fine-grained operations of HFFC, further indicating that the module leverages frequency-domain information to extract and enhance high-frequency features, rather than relying solely on spatial saliency. This mechanism effectively strengthens the representation of retinal edges that are challenging to capture in the spatial domain.

**Fig. 18** and **Fig. 19** show the results when dynamic decomposition and feature map selection with fusion are applied independently. In both cases, the feature maps contain considerable noise intermingled with edge information, which diminishes the model's focus on relevant details and consequently reduces detection accuracy. These observations underscore the critical importance of integrating dynamic decomposition with local selection and global fusion to achieve optimal feature enhancement and effective noise suppression.

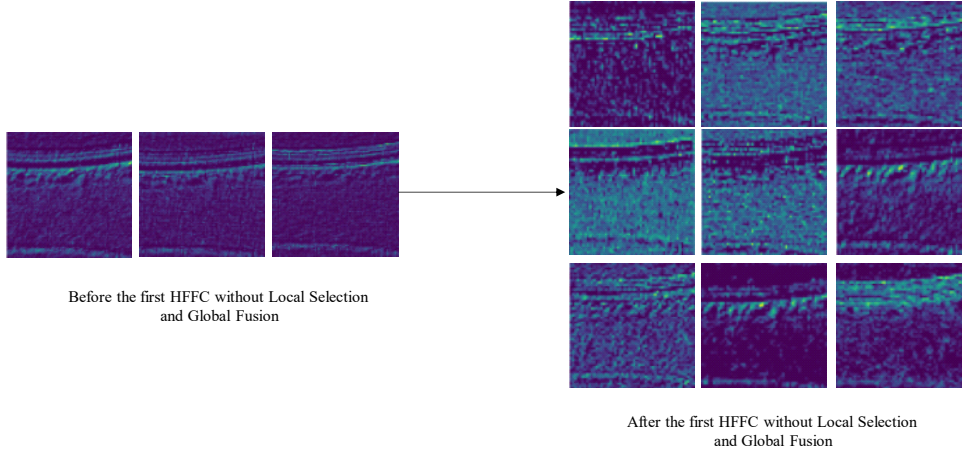

Before the first HFFC without Local Selection and Global Fusion

After the first HFFC without Local Selection and Global Fusion

**Fig. 18** Visualization of feature maps of the first HFFC without Local Selection and Global Fusion.

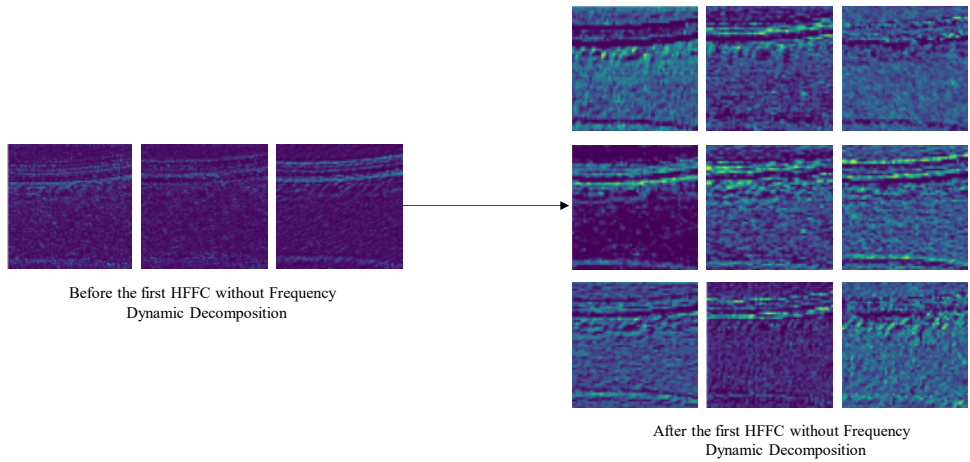

Before the first HFFC without Frequency Dynamic Decomposition

After the first HFFC without Frequency Dynamic Decomposition

**Fig. 19** Visualization of feature maps of the first HFFC without Frequency Dynamic Decomposition

**5.3 Robustness of WaveNet-SF to Noise**

    OCT images are often plagued by significant speckle noise, which can vary greatly depending on the medical device used for image acquisition. Some devices produce images with much higher noise levels, making robustness to noise a critical factor for real-world applications. While several existing models have demonstrated strong performance on standard datasets, many fail to account for the influence of speckle noise, resulting in poor performance when processing images with high noise levels.

    We observed that while multi-scale feature extraction methods can enhance model detection capabilities to some extent, their resistance to noise is notably reduced. For instance, when the noise intensity is 19.18dB, ResNet50's detection accuracy on the OCT-C8 and OCT2017 datasets drops by 16.67% and 5.68%, respectively, while FPN-ResNet50 suffers a decline of 35.22% and 8.02%. This deterioration can be attributed to the fact that larger-scale feature maps, while providing useful shallow semantic information and enhancing detection ability, also introduce noise when the noise level is high. As a result, these shallow feature maps significantly impair the model's noise resistance.

    In contrast, our WaveNet-SF model exhibits remarkable robustness to noise. First, the low-frequency components obtained through wavelet transform retain the global structural information of the image and perform preliminary denoising. The MSW-SA module, a multi-scale spatial attention mechanism, enables the model to focus on complex lesions of various sizes without relying on traditional

multi-scale feature extraction methods. Additionally, the lightweight HFFC module selectively extracts essential edge information from the noisy high-frequency components without introducing further noise, effectively enhancing the model's performance in noisy environments.

Despite the overall strong results in this paper, we observe that some categories show notably lower classification accuracy compared to others. This discrepancy may arise from the high inter-class similarity of lesions within these categories. To address this, future work could explore incorporating contrastive learning into the model to reduce intra-class distances and increase inter-class distances, thereby further enhancing the model's performance.

## 6. Conclusion

In this study, we propose the WaveNet-SF model, a hybrid learning approach that integrates wavelet transforms to process both spatial and frequency domains. The model decomposes images into high-frequency and low-frequency components, where the low-frequency components preserve global structural information. Through the MSW-SA module, the model can precisely focus on complex lesion areas within the low-frequency components. Meanwhile, the HFFC module enhances edge information from the high-frequency components, improving the model's ability to detect fine details. Ablation experiments validate the effectiveness of each component. Our model achieves state-of-the-art (SOTA) performance with accuracies of 97.82% and 99.58% on the OCT-C8 and OCT2017 datasets, respectively, and demonstrates excellent performance even in noisy environments. We believe that WaveNet-SF can provide valuable support for ophthalmologists in diagnostic tasks.

**Declaration of competing interest**

The authors declare that there are no conflicts of interest regarding the publication of this paper.

**CRediT authorship contribution statement**

**Jilan Cheng**: Conceptualization; Data curation; Formal analysis; Investigation; Methodology; Software; Validation; Visualization; Writing-original draft. **Guoli Long**: Investigation; Validation; Visualization; Writing-review & editing. **Zeyu Zhang**: Investigation; Validation; Writing-review & editing. **Zhenjia Qi**: Validation; Writing-review & editing. **Hanyu Wang**: Visualization; Writing-review & editing. **Libin Lu**: Visualization; Writing-review & editing. **Shuihua Wang**: Validation, Writing-review & editing. **Yudong Zhang**: Validation, Writing-review & editing. **Jin Hong**: Conceptualization; Data curation; Investigation; Methodology; Resources; Funding acquisition; Project administration; Supervision; Writing-review & editing.

**Acknowledgements**

This work was supported in part by the National Natural Science Foundation of China (62466033), and in part by the Jiangxi Provincial Natural Science Foundation (20242BAB20070).